\documentclass[12pt]{iopart}
\usepackage{graphicx}
\usepackage{color,soul}
\begin{document}

\title[]{Instability of spiral and scroll waves in the presence of a gradient
in the fibroblast density: the effects of fibroblast-myocyte coupling}

\author{Soling Zimik and Rahul Pandit}

\address{Centre for Condensed Matter Theory, Department of Physics, 
Indian Institute of Science, Bangalore, 560012, India}
\ead{rahul@physics.iisc.ernet.in}
\vspace{10pt}
\begin{indented}
\item[]September 2016
\end{indented}

\begin{abstract}

Fibroblast-myocyte coupling can modulate electrical-wave dynamics in cardiac
tissue. In diseased hearts, the distribution of fibroblasts is heterogeneous,
so there can be gradients in the fibroblast density (henceforth we call this
GFD) especially from highly injured regions, like infarcted or ischemic zones,
to less-wounded regions of the tissue. Fibrotic hearts are known to be prone to
arrhythmias, so it is important to understand the effects of GFD in the
formation and sustenance of arrhythmic re-entrant waves, like spiral or scroll
waves. Therefore, we investigate the effects of GFD on the stability of spiral
and scroll waves of electrical activation in a state-of-the-art mathematical
model for cardiac tissue in which we also include fibroblasts. By introducing
GFD in controlled ways, we show that spiral and scroll waves can be unstable in
the presence of GFDs because of regions with varying spiral or scroll-wave
frequency $\omega$, induced by the GFD. We examine the effects of the resting
membrane potential of the fibroblast and the number of fibroblasts attached to
the myocytes on the stability of these waves. Finally, we show that the
presence of GFDs can lead to the formation of spiral waves at high-frequency
pacing.

\end{abstract}

%
%
%
%
%

\section{Introduction}

The mechanical contractions of the heart muscles are mediated by electrical
waves of activation in cardiac tissue. Disturbances in the normal propagation
of these  electrical waves can be arryhthmogenic because of the excitation of
pathological re-entrant waves, such as spiral waves, in two dimensions (2D),
and scroll waves, in three dimensions (3D).  Spiral waves  are linked to
cardiac arrhythmias, such as ventricular tachycardia (VT) and  ventricular
fibrillation (VF). Ventricular tachycardia is known to be associated with the
existence of a single spiral (or scroll) wave in the
medium~\cite{efimov1999evidence,de1988reentry}; life-threatening VF is
associated with multiple or broken spiral (or scroll)
waves~\cite{bayly1998spatial,witkowski1998spatiotemporal,walcott2002endocardial}.
Some studies indicate that episodes of VF are preceded by
VT~\cite{morita2009increased,makikallio1999heart}, suggesting the possibility
of an initial onset of a spiral wave, which then degenerates to a
multiple-spiral-wave turbulent state
(VF)~\cite{courtemanche1996complex,karma1993spiral}. Given that VF can prove to
be fatal, it is important to understand this transition from VT to VF. Many
factors can affect the stability of a spiral wave; hence multiple
mechanisms have been proposed for this VT-VF
transition~\cite{fenton2002multiple}. Here, we discuss the effects of
inhomogeneities in the fibroblast density distribution on the stability of
spiral and scroll waves of electrical activation in a state-of-the-art
mathematical model for cardiac tissue, in which we also include fibroblasts.

Fibroblast proliferation (fibrosis) occurs in the heart during myocardial
remodelling in, e.g., ischemic hearts or hypertensive
hearts~\cite{manabe2002gene}; such proliferation is part of the wound-healing
process after injuries caused, e.g., by myocardial infarction. When fibroblasts
are coupled to a cardiac myocyte, they can change the electrophysiological
properties of the myocyte
\cite{xie2009cardiac,nguyen2012arrhythmogenic,sachse2009model,maccannell2007mathematical}.
This modulation of the electrophysiological properties of myocytes can, in
turn, alter the dynamics of waves of electrical activation in cardiac tissue.
Also, because of the heterocellular coupling between fibroblasts and myocytes,
the presence of fibroblasts modulates the conduction properties of cardiac
tissue.  Therefore, the interposition of fibroblasts between myocytes in
cardiac tissue can lead to the fragmentation of the electrical
waves~\cite{zlochiver2008electrotonic,majumder2012nonequilibrium} and even
waveblock \cite{majumder2012nonequilibrium}. Many studies have investigated the
effects of fibroblasts on wave dynamics in cardiac
tissue~\cite{zlochiver2008electrotonic,majumder2012nonequilibrium,nayak2013spiral,ten2007influence,kazbanov2016effects}.
Some of these studies model the fibroblasts as inexcitable obstacles
\cite{ten2007influence,kazbanov2016effects}; others take into account the
fibroblast-myocyte coupling and consider either (a) a random distribution of
fibroblasts, with an average density that is uniform in
space~\cite{zlochiver2008electrotonic,majumder2012nonequilibrium,nayak2013spiral},
or (b) localized fibroblast inhomogeneities~\cite{nayak2013spiral}. However, in
real diseased hearts the distribution of fibroblasts may not be uniform, even
on average, but, rather, there may be a gradient in fibroblast density (GFD),
as has been observed in aged-rabbit hearts~\cite{morita2009increased}.
Moreover, in hearts that have been injured, say because of myocardial infarction,
the fibroblast density may vary from a high value in the infarcted region to a
lower value in the normal region of the
heart~\cite{nguyen2014cardiac,ashikaga2013feasibility}, with intermediate
values in interfaces between these regions. It is important, therefore, to
understand what role such GFD can play in inducing and then, perhaps,
destabilizing re-entrant waves, like spiral or scroll waves. 

We show that a state-of-the-art mathematical model for cardiac tissue, based
on the O'Hara-Rudy model (ORd) for a human ventricular
cell~\cite{o2011simulation}, provides us with a natural platform for (a)
incorporating fibroblast-myocyte interactions and (b) imposing GFD in a
controlled way so that we can study, \textit{exclusively}, its effects on spiral- and
scroll-wave dynamics, without other complicating factors that can be present in
real cardiac tissue, such as scars, which lead to conduction
inhomogeneities~\cite{shajahan2009spiral}.  We carry out such a controlled
study of the effects of GFD by using the ORd model, for a human ventricular
cell \cite{o2011simulation}, with passive fibroblasts, as in the model of
MacCannell, \textit{et al.}~\cite{maccannell2007mathematical}, which interact
with myocytes (see Materials and Methods).  We introduce fibroblasts in such a
way that we can control GFD systematically.  Before we present the details of
our study, we give a qualitative overview of our principal results. 

We find that GFD can destabilize spiral or scroll waves, i.e., lead to the
break up of a spiral or scroll wave into multiple waves. Our \textit{in silico}
studies help us to uncover the principal mechanism of this GFD-induced spiral-
or scroll-wave instability. The myocyte-fibroblast coupling changes the action
potential duration (APD) of the attached myocytes. We show first how the
spiral-wave frequency $\omega$, in a homogeneous domain with randomly
distributed fibroblasts, depends inversely on the APD of the myocytes in the
medium, a result that is consistent with  dimensional analysis. Roughly
speaking, GFD induces a gradient in the mean APD, i.e., $\overline{\rm{APD}}$, in the
medium; therefore, if a spiral wave forms, with its core in a
high-$\overline{\rm{APD}}$ region, its rotation frequency near the core is low. Such a
low-$\omega$ region cannot support wave trains that come from a high-$\omega$
region, say because of a spiral wave there. We find that the greater the
variation of $\overline{\rm{APD}}$ with GFD the more readily does spiral- or
scroll-wave instability set in.  We build on this qualitatively appealing
argument by carrying out detailed numerical investigations of GFD-induced
spiral- and scroll-wave break up, in two- and three-dimensional simulation
domains, respectively, and in three models, the first with a uniform
distribution of fibroblasts, and two others in which the mean density of
fibroblasts changes either (a) linearly along one spatial direction or (b) as a
step function.  Furthermore, we investigate the effects of other fibroblast
parameters, like the resting membrane potential $\rm{E_f}$ and the number
$\rm{N_f}$  of fibroblasts attached to myocytes, on the stability of such
spiral and scroll waves. For a given GFD, we find that the larger the values of
$\rm{E_f}$ and $\rm{N_f}$, the more readily does spiral- or scroll-wave
instability set in. Finally, we show how the presence
of GFD can lead to the formation of re-entrant waves, like spiral waves, when
we pace an edge of our simulation domain at a  high frequency.

The remaining part of this paper is organized as follows. The {\bf Materials
and Methods} Section describes the models we use and the numerical methods we
employ to study them. The Section with the title {\bf Results} contains our
results, from single-cell and tissue-level simulations. Finally, the Section
called {\bf Discussions} is devoted  to a discussion of our results in the
context of earlier numerical and experimental studies.

\section{Materials and Methods}

We use the state-of-the-art O'Hara-Rudy model (ORd) for a human ventricular
cell \cite{o2011simulation} for our myocyte cell. In the ORd model, we
implement the modifications suggested in Ref.~\cite{o2011simulation}, where the
fast sodium current ($\rm{I_{Na}}$), of the original ORd model, has been
replaced with that of the model due to Ten Tusscher and Panfilov
\cite{ten2006alternans}. The fibroblasts are modelled as passive cells, for
which we use the model given by MacCannell, \textit{et al.}
\cite{maccannell2007mathematical}, but, instead of using a constant membrane
conductance $\rm{G_f}$, we use a $\rm{G_f}$ that has a nonlinear dependence on
its membrane voltage $\rm{V_f}$, as in Refs. \cite{zlochiver2008electrotonic,
rook1992differences}. The value of $\rm{G_{f}}$ is set to 1 nS if $\rm{V_f}$ is
below -20 mV, and 2 nS for $\rm{V_f}$ above -20 mV. The gap-junctional
conductance between a myocyte and a fibroblast is set to 8 nS. 

\begin{figure}[!ht]
\includegraphics[width=\linewidth]{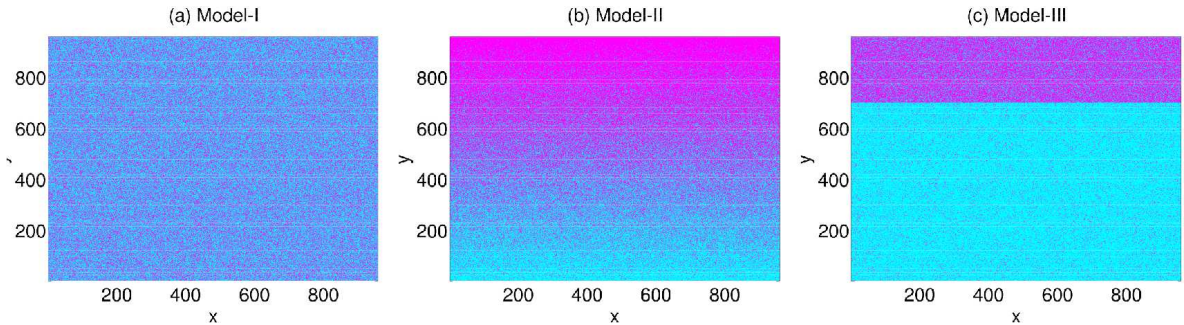}
\caption{ { \bf Models of fibroblast distribution.} Pseudocolor plots showing
our three models of fibroblast distribution. (a) Model-I: the density
distribution of fibroblasts is uniform throughout the domain, on average. The
cyan color indicates myocyte density with no fibroblasts, and the magenta color
indicates fibroblast-myocyte composites. The value of $\rm{p_f}$ is 40\%. (b)
Model-II: there is a gradient of fibroblast density (GFD) along the vertical y
axis, with $\rm{p_{f1}}$= 10\%  and $\rm{p_{f2}}$= 100\% (see
equation~\ref{model2}). (c) Model-III: $\rm{p_f}$ changes abruptly from 10\%,
in the bottom $\frac{3}{4}$ of the domain, to 70\% in the upper $\frac{1}{4}$
of the domain (see equation~\ref{model3}). } \label{model}
\end{figure}

In our two-dimensional (2D) simulations the fibroblasts are attached atop the
myocytes; thus, our 2D simulation domain is a bilayer, as in previous studies
\cite{nayak2013spiral,zimik2015computational}; similarly, we place fibroblasts
in the interstitial region between myocytes in our three-dimensional (3D)
simulations as in~Ref.~\cite{nayak2015turbulent}. The number of fibroblasts
$\rm{N_f}$ attached to a myocyte in a fibroblast-myocyte composite is 2, unless
otherwise mentioned in the text. For a given percentage $\rm{P_f}$ of
fibroblasts in the medium, the fibroblasts are attached to the myocytes in a
tissue with a probability $\rm{P_f}/100$.  We carry out three types
of simulations for three different models of fibroblast distribution: Model-I,
Model-II, and Model-III. In Model I, the value of $\rm{P_f(x,y)}$ is constant
throughout the domain, as in equation~(\ref{model1}) (see figure~\ref{model}
(a)). Thus, the distribution of fibroblast is uniform and isotropic.  In Model
II, the value of $\rm{P_f(x,y)}$ varies linearly along the vertical direction
(y axis) of the domain (see figure~\ref{model} (b)) given by
equation~(\ref{model2}). In the last Model III, the value of $\rm{P_f(x,y)}$
changes discontinuously (see figure~\ref{model} (c)) given by
equation~(\ref{model3}).

\begin{equation}
\rm{Model~I:}~~~~\rm{P_f(x,y)} = \rm{p_f}~\forall~\rm{(x,y)},
\label{model1} 
\end{equation}    
where $\rm{p_f}$ is a constant between 0 and 100.

\begin{equation}
\rm{Model~II:}~~~~\rm{P_f(x,y)} =  \rm{p_{f1}}+\frac{y(\rm{p_{f2}}-\rm{p_{f1}})}{L}, 0\leq \rm{y} \leq \rm{L},
\label{model2}
\end{equation}
where L is the size of the square domain, and $\rm{p_{f1}}$ and $\rm{p_{f2}}$
are constants between 0 and 100.
 
\begin{equation}
\rm{Model~III:}~~~~\rm{P_f(x,y)} = \rm{p_{f1}}\times\Theta(y_0-y) + \rm{p_{f2}}\times\Theta(y-y_0) ,
\label{model3}
\end{equation}
where $\Theta(\rm{y})$ is the Heaviside step function, and $\rm{y_0}$ is a
particular value of the distance along the vertical direction; we use
$\rm{y_0}$ = $\frac{3}{4}$L in our study.

The membrane potential of a myocyte $\rm{V_m}$ is governed by the ordinary
differential equation (ODE)
\begin{equation}
\frac{{\rm{d V_m}}}{\rm{d t}} = -\frac{I_{\rm{ion}}+\rm{N_f} \times I_{\rm{gap}}}{C_m},\label{ode1}
\end{equation}
where $C_m$ is the myocyte capacitance, which has a value of 185 pF;
${I_{\rm{ion}}}$ is the sum of all the ionic currents of the myocyte,
${I_{\rm{gap}}}$ is the gap-juctional current between the fibroblast and the
myocyte, and $\rm{N_f}$ is the number of fibroblasts attached to the myocyte.
We give ${I_{\rm{ion}}}$ and ${I_{\rm{gap}}}$ below:
	
\begin{equation}
I_{\rm{ion}}= \rm{I_{Na}}+ \rm{I_{to}}+ \rm{I_{CaL}}+ \rm{I_{CaNa}}+ \rm{I_{CaK}}+ \rm{I_{Kr}}+ \rm{I_{Ks}}+ \rm{I_{K1}}+ \rm{I_{NaCa}}+ \rm{I_{NaK}}+ \rm{I_{Nab}}+
			\rm{I_{Cab}}+ \rm{I_{Kb}}+ \rm{I_{pCa}};
\end{equation}

\begin{equation}
I_{\rm{gap}}= \rm{G_{gap}}(\rm{V_m}-\rm{V_f});
\end{equation}

we list the ionic currents of the myocyte in Table~\ref{table1}.

The membrane potential of a fibroblast is governed by the equation
\begin{equation}
\frac{{\rm{d V_f}}}{\rm{d t}} = \frac{I_{\rm{gap}}-I_{\rm{f}}}{C_f},
\label{ode2}
\end{equation}
where ${C_f}$=6.3 pF is the membrane capacitance of the fibroblast, and $I_{\rm{f}}$ is the fibroblast current,

\begin{equation}
I_{\rm{f}}= \rm{G_{f}}(\rm{V_f}-\rm{E_f});
\end{equation}

here $\rm{E_f}$ is the resting membrane potential of the fibroblast, and its
range of values in our study is from 0 to -50 mV, as observed experimentally in
Refs.~\cite{rook1992differences,camelliti2005structural, kamkin1999mechanically,
kohl2003heterogeneous}.

\begin{table}[!ht]
\caption{{\bf Table of currents.}}
	\begin{tabular}{l l }
	\multicolumn{2}{l}{} \\
	\hline
	\hline
	$\rm{I_{Na}}$ & fast inward $\rm{Na^+}$ current \\
	$\rm{I_{to}}$ & transient outward $\rm{K^+}$ current \\
	$\rm{I_{CaL}}$ & L-type $\rm{Ca^{2+}}$ current \\
	$\rm{I_{Kr}}$ & rapid delayed rectifier $\rm{K^+}$ current \\
	$\rm{I_{Ks}}$ & slow delayed rectifier $\rm{K^+}$ current \\
	$\rm{I_{K1}}$ & inward rectifier $\rm{K^+}$ current \\
	$\rm{I_{NaCa}}$ & $\rm{Na^+/Ca^{2+}}$ exchange current \\
	$\rm{I_{NaK}}$ & $\rm{Na^+/K^+}$ ATPase current\\
	$\rm{I_{Nab}}$ & $\rm{Na^+}$ background current \\
	$\rm{I_{Cab}}$ & $\rm{Ca^{2+}}$ background current\\
	$\rm{I_{pCa}}$ & sarcolemmal $\rm{Ca^{2+}}$ pump current\\
	$\rm{I_{Kb}}$ & $\rm{K^+}$ background current\\
	$\rm{I_{CaNa}}$ & $\rm{Na^+}$ current through the L-type $\rm{Ca^{2+}}$ channel \\
	$\rm{I_{CaK}}$ & $\rm{K^+}$ current through the L-type $\rm{Ca^{2+}}$ channel\\
	\hline
	\end{tabular}
	\begin{flushleft} A list of the ionic currents in the ORd model (symbols as in Ref.~\cite{o2011simulation}).
		 \end{flushleft}
\label{table1}
\end{table}

The spatiotemporal evolution of the membrane potential ($\rm{V_m}$) of the
myocytes in tissue is governed by a reaction-diffusion equation, which is the
following partial-differential equation (PDE):

\begin{equation}
\frac{\partial{\rm{V_m}}}{\partial{t}}+\frac{I_{\rm{ion}} +\rm{N_f}\times I_{\rm{gap}}}{C_m}=\rm{D}\nabla^2 \rm{V_m},
\label{pde}
\end{equation} 
where $\rm{D}$  is the diffusion constant between the myocytes.

\subsection{Numerical Methods}

\begin{figure}[!ht]
\includegraphics[width=\linewidth]{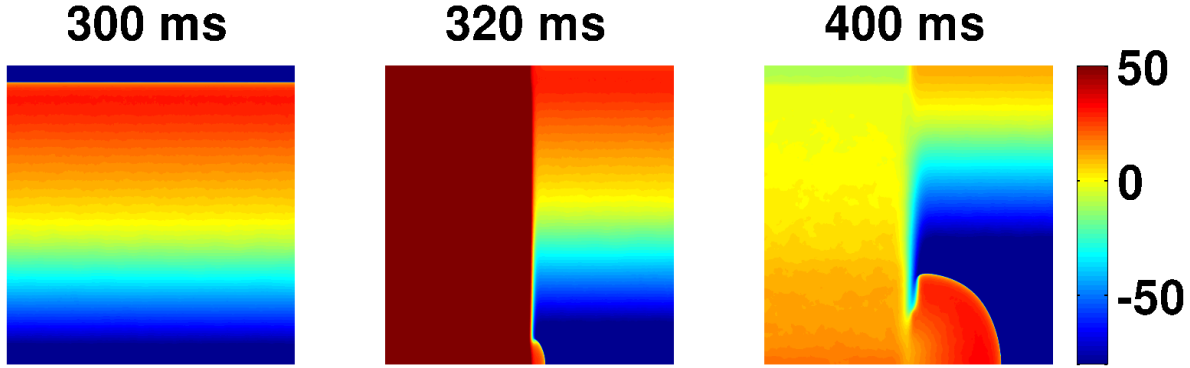}
\caption{ { \bf S1-S2 cross-field protocol.} Pseudocolor plot illustrating the
S1-S2 cross-field protocol that we use to initiate a spiral wave in the domain.
The colorbar indicates the membrane voltage $\rm{V_m}$ in millivolts.}
\label{protocol}
\end{figure}

We solve the ODEs~(\ref{ode1}) and (\ref{ode2}) for $\rm{V_m}$ and $\rm{V_f}$,
respectively, and also the ODEs for the gating variables of the ionic currents
of the myocyte by using a forward-Euler method. For solving the
PDE~(\ref{pde}), we use the forward-Euler method for time marching with a
five-point stencil for the Laplacian in 2D and a 7-point stencil in 3D. We set
D= 0.0012 $\rm{cm^{2}/ms}$. The temporal and spatial resolutions are set to be
$\delta{x}$= 0.02 cm and $\delta{t}$=0.02 ms, respectively. The conduction
velocity of a plane wave in the tissue, with the above set of parameters, is 65
cm/s. In our two-dimensional (2D) tissue simulations, we use a domain size of
$\rm{960\times960}$ grid points, which  translates into a physical size of
$\rm{19.2\times19.2 cm^2}$. We initiate the spiral wave by using the
conventional S1-S2 cross-field protocol, where we first apply an S1 plane wave
and allow its wave back to cross some part of the domain (see
figure~\ref{protocol}, time= 300 ms) and then we apply the S2 stimulus
perpendicular to the S1 wave (figure~\ref{protocol}, time= 320 ms), as shown in
figure~\ref{protocol}. The strength and duration of both the S1 and S2 stimuli
are -150 $\mu\rm{A}/\mu\rm{F}$ and $3$~ms, respectively. All our tissue
simulations are carried out for a duration of 10 seconds.
  
\section{Results}

We first study the effects of the gap-junctional coupling $\rm{G_{gap}}$ of
fibroblasts on the electrophysiological properties of myocytes. We then
investigate the effects of $\rm{G_{gap}}$  on the spiral-wave frequency in a 2D
simulation. Next we show how GFD leads to spiral-wave instability. Thereafter,
we show the effects of GFD on scroll-wave stability in a 3D simulation domain.
Finally, we discuss the role of GFD on the initiation, via pacing, of
re-entrant spiral waves.

\subsection{Effects of fibroblast gap-junctional coupling, on a myocyte, and
the spiral-wave frequency}

\begin{figure}[!ht]
\includegraphics[width=\linewidth]{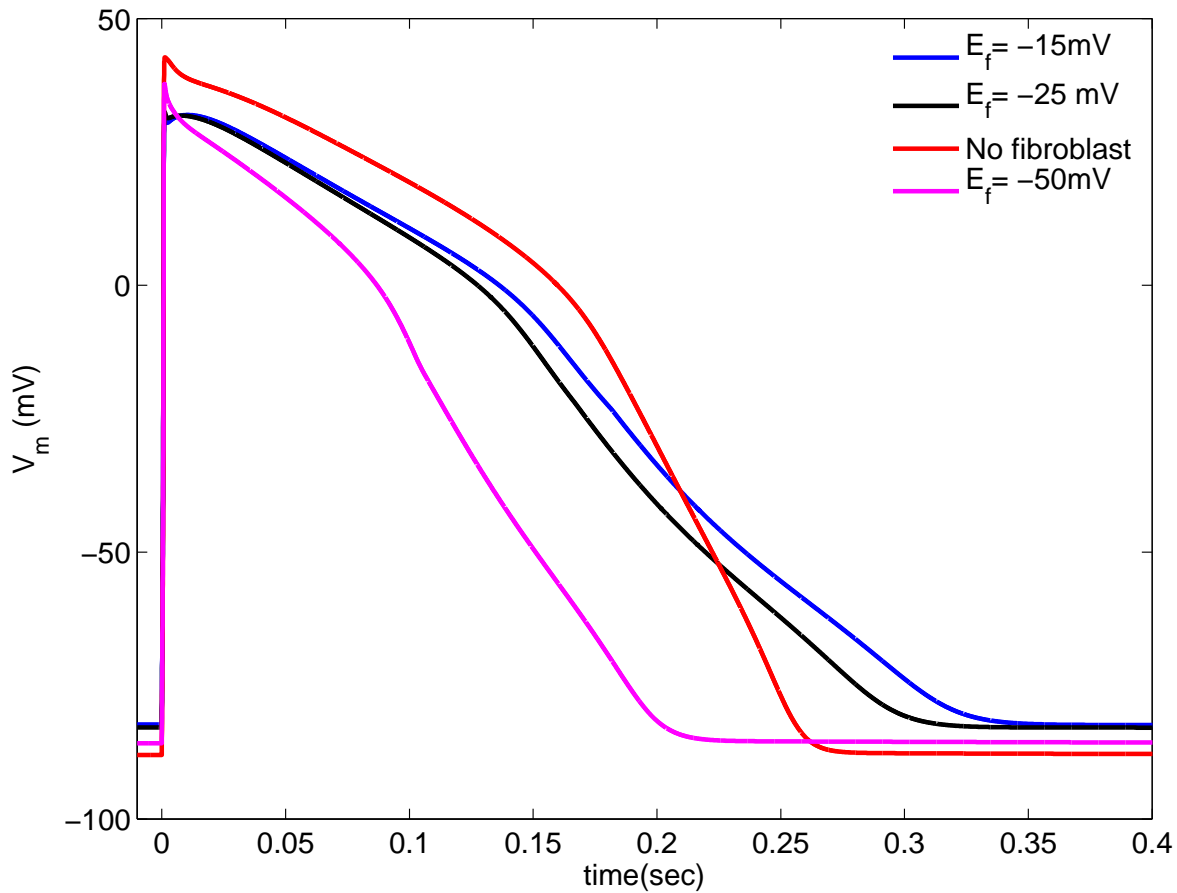}
\caption{ { \bf Action potentials of an isolated myocyte and a myocyte
attached to fibroblasts.} Plots showing the action potentials of an isolated
myocyte (red curve), and a myocyte attached to $\rm{N_f}$= 2 fibroblasts of
$\rm{E_f}$= -15 mV (blue curve), and $\rm{E_f}$= -25 mV (black curve), and
$\rm{E_f}$= -50 mV (magenta curve). The APDs of the myocyte attached to
fibroblasts of $\rm{E_f}$ = -15 mV and $\rm{E_f}$= -25 mV are larger than that of
the isolated myocyte, whereas the APD of the myocyte attached to fibroblasts of
$\rm{E_f}$ =-50 mV is lower than that of the isolated myocyte. } \label{AP}
\end{figure}

The gap-junctional coupling of fibroblasts with a myocyte changes the
electrophysiological properties of the myocyte
\cite{xie2009cardiac,nguyen2012arrhythmogenic,sachse2009model,maccannell2007mathematical}.
For instance, it can modulate the action potential duration (APD) of a myocyte.
The APD of the myocyte may increase or decrease depending on the
electrophysiological properties of the fibroblasts, like their resting-membrane
potential $\rm{E_f}$. Figure~\ref{AP} shows the action potentials of a myocyte
attached to $\rm{N_f}$= 2 fibroblasts with different values of $\rm{E_f}$. The
APDs of the myocyte attached to fibroblasts, with $\rm{E_f}$= -15 mV (blue
curve) and $\rm{E_f}$= -25 mV (black curve) are larger than that of an isolated
myocyte (red curve), whereas the APD of the myocyte attached to fibroblasts
with $\rm{E_f}$= -50 mV (magenta curve) is lower than that of an isolated
myocyte. 

This fibroblast-induced modulation of the APDs of the myocytes affects the
properties of electrical waves, like the spiral-wave frequency, at the tissue
level. This spiral frequency $\omega$ depends on the APDs of the constituent
myocytes as follows. Consider a stable spiral that does not meander;
dimensional analysis yields

\begin{equation} 
\omega \simeq \frac{\theta}{\lambda}, 
\end{equation} 

where $\theta$ is the conduction velocity and $\lambda$ is the wavelength. 
Furthermore, $\lambda$ $\simeq$ $\theta\times\rm{APD}$, and, therefore, 
\begin{equation} 
\omega\simeq\frac{1}{\rm{APD}}. 
\label{omega_apd} 
\end{equation}

\begin{figure}[!ht]
\includegraphics[width=\linewidth]{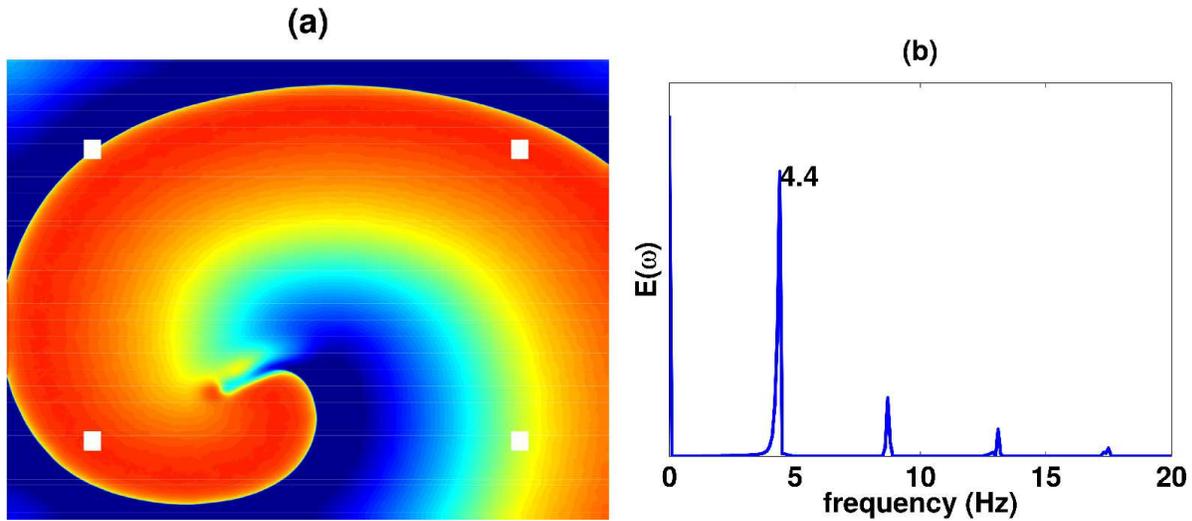}
\caption{ { \bf A spiral wave in a domain with a uniform and isotropic
distribution of fibroblasts.} a) The spiral wave supported in the domain with
$\rm{p_f}$= 40\%.  b) The averaged power spectrum of the time-series recording
of $\rm{V_m}$ from four representative points, indicated by white squares in
(a), located near the four corners of the simulation domain. The spiral
frequency $\omega$ is determined by considering the dominant peak, which is 4.4
Hz in this case; the other major peaks are harmonics at 8.8, 13.2, and 17.6 Hz.} \label{Pf40}
\end{figure}

\begin{figure}[!ht]
\includegraphics[width=\linewidth]{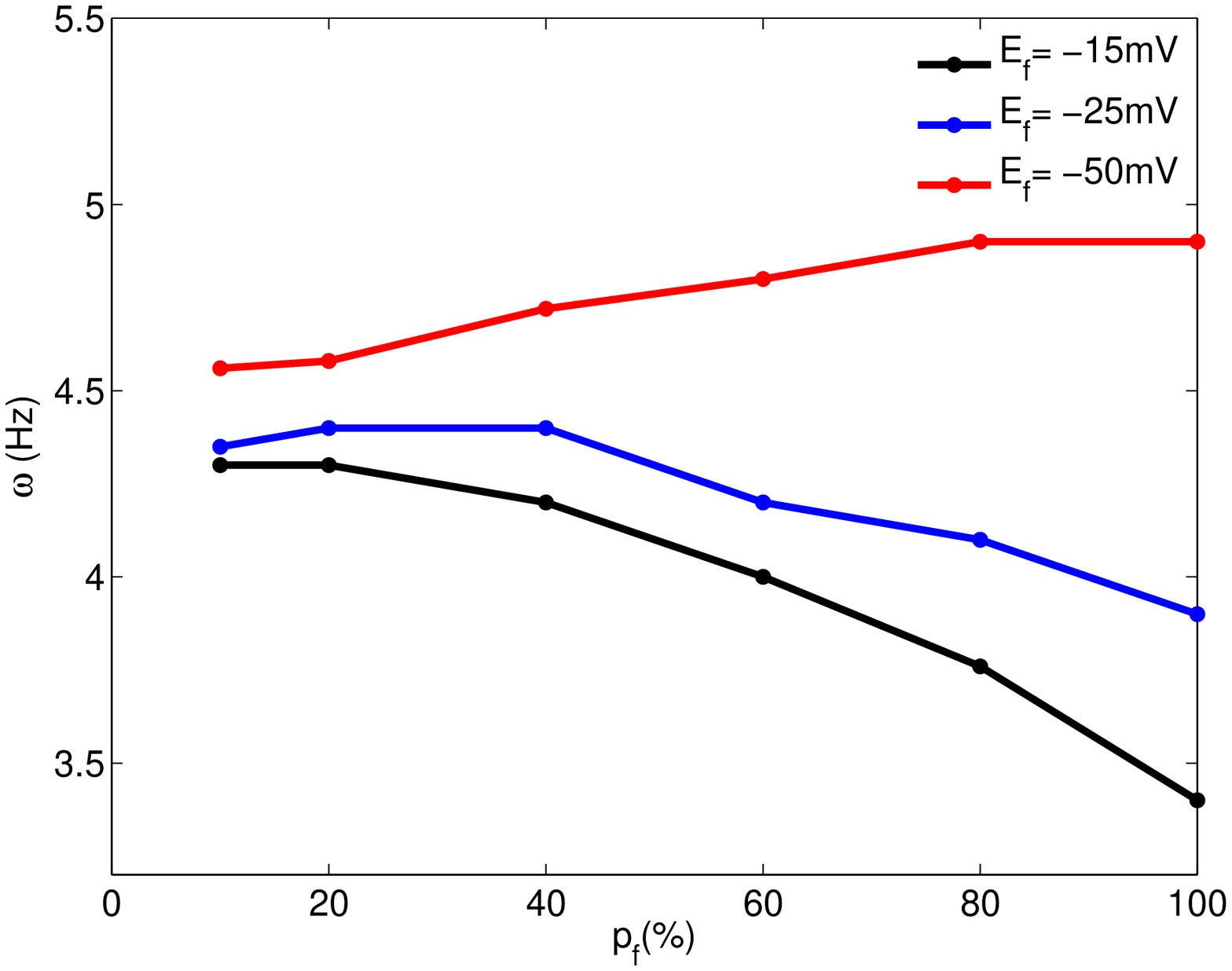}
\caption{ { \bf Variation of the spiral-wave frequency $\omega$ with the
percentage of fibroblasts $\rm{p_f}$, in a domain with a uniform and isotropic
fibroblast distribution (Model-I), for three different values of $\rm{E_f}$.}
Plots showing the variation of  $\omega$ with $\rm{p_f}$ for $\rm{E_f}$= -15 mV
(black curve), $\rm{E_f}$= -25 mV (blue curve), and $\rm{E_f}$= -50 mV (red
curve). $\omega$ decreases with $\rm{p_f}$ for $\rm{E_f}$= -15 mV and -25 mV,
but increases for $\rm{E_f}$= -50 mV. The variation of $\omega$ with $\rm{p_f}$
is most significant for $\rm{E_f}$= -15 mV.} \label{freq}
\end{figure}

In a domain with a uniform and isotropic fibroblast distribution with
$\rm{P_f(x,y)}$=$\rm{p_f}$ (Model-I, equation~\ref{model1} and
figure~\ref{model} (a)), the average APD ($\overline{\rm{APD}}$) of the
myocytes depends on $\rm{p_f}$, and, therefore, the frequency $\omega$ of a
spiral wave depends on $\rm{p_f}$. In figure~\ref{freq} we plot $\omega$ versus
$\rm{p_f}$ for $\rm{E_f}$= -50mv (red curve), and $\rm{E_f}$= -25 mV (blue
curve), and $\rm{E_f}$= -15 mV (black curve). We see from the plot for
$\rm{E_f}$= -15 mV that $\omega$ decreases as $\rm{p_f}$ increases; this
decrease is slower for $\rm{E_f}$= -25 mV; and, by the time $\rm{E_f}$= -50 mV,
$\omega$ increases with $\rm{p_f}$.  The decrease of $\omega$ with $\rm{p_f}$
for $\rm{E_f}$= -25 mV and -15 mV is because fibroblasts, with these values of
$\rm{E_f}$, increase the APD of an attached myocyte (see figure~\ref{AP}); and,
therefore, in a tissue with a uniform and isotropic fibroblast distribution,
the increase of $\rm{p_f}$ increases $\overline{\rm{APD}}$ of the constituent
myocytes, and, thereby, decreases $\omega$ (see equation~(\ref{omega_apd})).
However, for $\rm{E_f}$= -50 mV, the APD of the myocyte decreases (see
figure~\ref{AP}), so $\omega$ increases with $\rm{p_f}$. For the plots in
figure~\ref{freq} we obtain the spiral frequency $\omega$ from the averaged
power spectra of time-series of $\rm{V_m}$, which we record from four
representative points of the domain (white squares in figure~\ref{Pf40} (a)).
We show one representative example for $\rm{p_f}$= 40\% in figure~\ref{Pf40}.
Figure~\ref{Pf40}(a) shows the spiral wave in the domain; and
figure~\ref{Pf40}(b) shows the power spectrum of $\rm{V_m}$ averaged over those
from the points (white squares in figure~\ref{Pf40}(a)) of the domain. The
spiral-wave frequency $\omega$ is the value of the dominant peak in the
spectrum, which is 4.4 Hz in this case.

\subsection{Instability of spiral waves}

\begin{figure}[!ht]
\includegraphics[width=\linewidth]{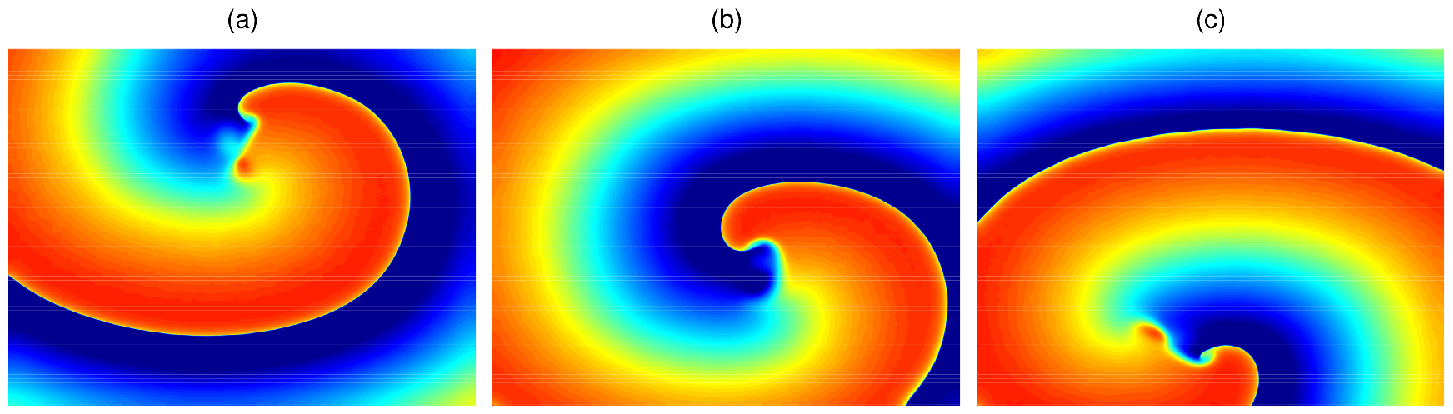}
\caption{ { \bf Variation of the spiral-wave frequency $\omega$, near the
spiral tip (see text), with space in a domain with GFD.} Pseudocolor plot of
$\rm{V_m}$ showing the three cases, where the spiral wave is initiated at the
top (a), middle (b), and bottom (c) parts of the domain, with a Model-II
fibroblast distribution, $\rm{p_{f1}}$= 10\%, $\rm{p_{f2}}$= 80\%, and
$\rm{E_f}$= -25 mV.  The spiral-wave frequency $\omega$ in the top, middle, and
bottom regions is 4.2 Hz, 4.3 Hz, and 4.4 Hz, respectively.} \label{TMB}
\end{figure}

\begin{figure}[!ht]
\includegraphics[width=\linewidth]{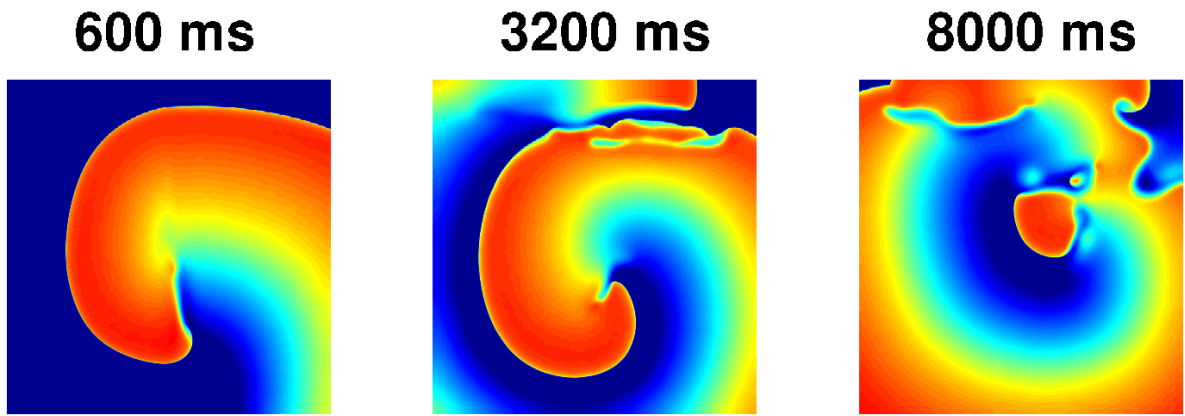}
\caption{ { \bf Spiral-wave instability in the presence of GFD.} Pseudocolor
plot of $\rm{V_m}$ showing the spatiotemporal evolution of the break-up of a
spiral in the presence of GFD, with the Model-II distribution
(figure~\ref{model} (b)), $\rm{p_{f1}}$= 10\%, $\rm{p_{f2}}$= 100\%, and
$\rm{E_f}=$ -25 mV. The spiral arm breaks up  in the
upper region, where the spiral frequency $\omega$ supported by the region is
lower than that in the bottom region. } \label{Ef25}
\end{figure}

The spiral-wave frequency $\omega$, in a domain with fibroblasts, depends on
the fibroblast density $\rm{p_f}$. Therefore, in a domain with an inhomogeneous
distribution of fibroblasts, the spiral-wave frequency (the frequency at which
the spiral tip rotates) varies with space. To illustrate this, we take the
Model-II fibroblast distribution (see equation~\ref{model2}), where
$\rm{P_f(x,y)}$ varies along the y axis. This variation of $\rm{P_f(x,y)}$ in
the y direction induces a y dependence in $\overline{\rm{APD}}$ and, therefore,
in $\omega$. We set $\rm{p_{f1}}$= 10\%, $\rm{p_{f2}}$= 80\%, and $\rm{E_f}$=
-25 mV. Figure~\ref{TMB} show the cases when a spiral wave is initiated in
three different regions of the domain: (a) top, (b) middle, and (c) bottom. The
frequencies $\omega$ of the spiral tip in these top, middle, and bottom
regions are 4.2 Hz, 4.3 Hz, and 4.4 Hz, respectively. We measure these
frequencies from the time series of $\rm{V_m}$, which we record
from the four representative points, shown in figure~\ref{Pf40} (a). The spatial dependence of the local value of
$\omega$ in the domain can lead to spiral-wave instability if the variation in
$\omega$ is sufficiently large. We show this for the Model-II fibroblast
distribution, with $\rm{p_{f1}}$= 10\%, $\rm{p_{f2}}$= 100\%, and $\rm{E_f}$=
-25 mV. If we initiate a spiral somewhere in the middle of our simulation
domain, it breaks up as shown in figure~\ref{Ef25} \textit{in the upper region
of the domain, where} $\omega$ \textit{is low} (see also the Supplementary Video S1). This can be understood
qualitatively as follows: The upper, low-$\omega$ region cannot support the
high-frequency wave-trains that are emitted by the spiral tip, which is
rotating in  the middle region, with a higher value of $\omega$ than in
the upper part of the domain; the inability of the upper region to support
high-frequency waves leads to anisotropic conduction blocks because of the
anisotropic fibroblast distribution and, hence, wave breaks.

This breaking of waves in the low-$\omega$ region (i.e., large
$\overline{\rm{APD}}$ region) has also been observed experimentally in
monolayers of neonatal-rat ventricular myocytes by Campbell, \textit{et al.}
\cite{campbell2012spatial}. In these experiments, the gradient in
$\overline{\rm{APD}}$ has been induced by varying the $\rm{I_{Kr}}$ ion-channel
density.

\begin{figure}[!ht]
\includegraphics[width=\linewidth]{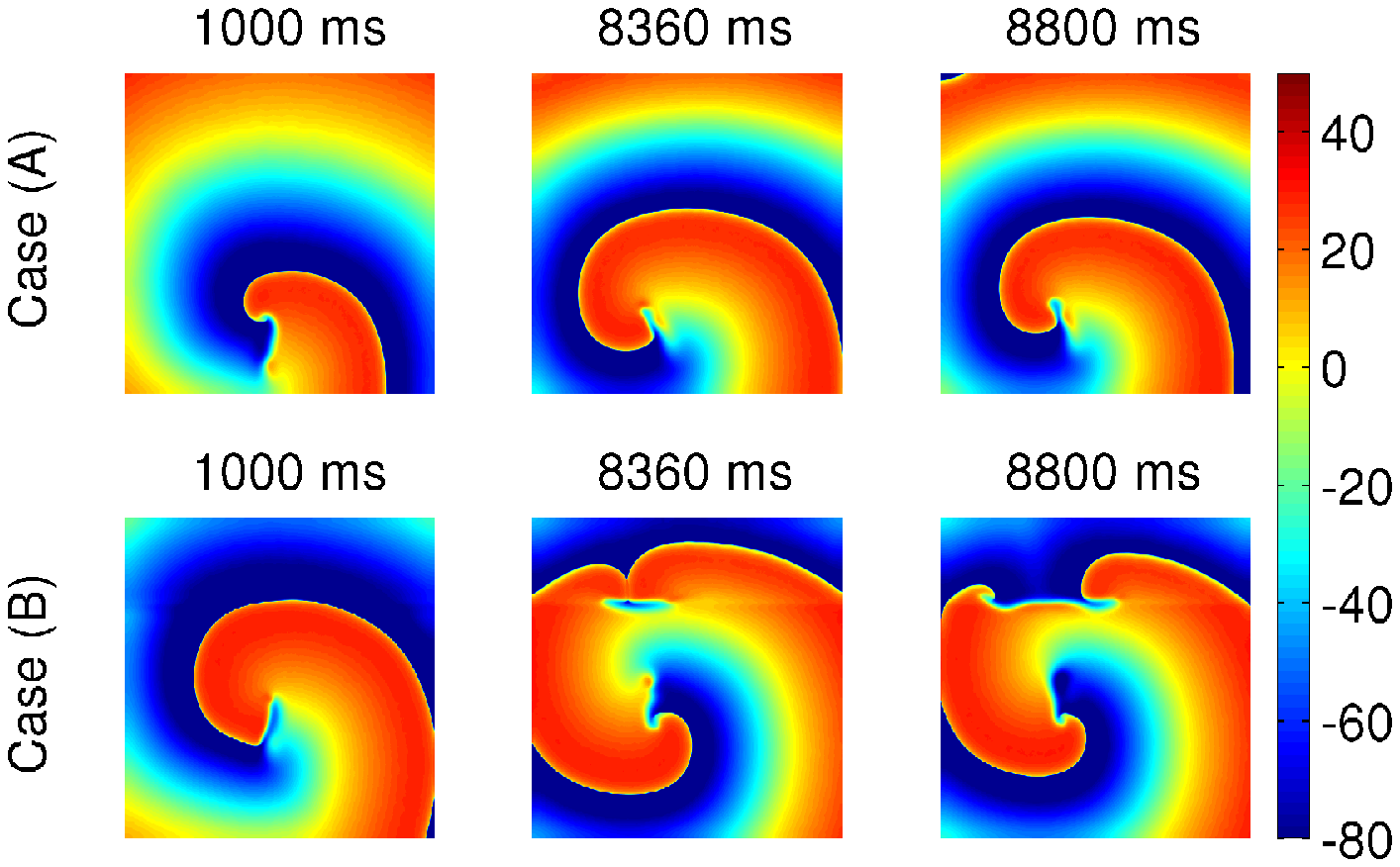}
\caption{ { \bf Dependence of spiral-wave stability on the value of the
gradient of $\omega$ induced by the GFD.} Pseudocolor plot of $\rm{V_m}$
showing the spatiotemporal evolution of a spiral for two cases (A) and (B) of
the fibroblast distribution. Top-panel figures show a stable spiral for case
(A), with the Model-II fibroblast distribution, $\rm{p_{f1}}$= 10\%, and
$\rm{p_{f2}}$= 75\%. The bottom panels shows spiral break-up for case (B), with
the Model-III fibroblast distribution, $\rm{p_{f1}}$= 10\%, and $\rm{p_{f2}}$=
65\% in equation~\ref{model3}.} \label{10and70}
\end{figure}

We have shown that spiral-wave instability stems from the spatial gradient in
$\omega$ that is induced by the GFD. We might expect, therefore, that steep
changes in $\omega$ might lead to an enhancement in such instability. To show
this, we study the following two cases: case (A), with the Model-II fibroblast
distribution, $\rm{p_{f1}}$= 10\%, and $\rm{p_{f2}}$= 75\%; and case (B), with
the Model-III fibroblast distribution, given in equation~\ref{model3} and
illustrated in figure~\ref{model} (c), with $\rm{p_{f1}}$= 10\% and
$\rm{p_{f2}}$= 65\%. The value of $\rm{E_f}$ for both these cases is -25 mV. We
show in figure~\ref{10and70} that the spiral breaks in case (B)
(figure~\ref{10and70} top panel), but not in case (A) (figure~\ref{10and70}
bottom panel), although the maximum fibroblast percentage is higher in case
(A), with the Model-II distribution, than in case (B),  with the Model-II
distribution (see also the Supplementary Video S2). Therefore, a comparison of our results for cases (A) and (B)
shows that a high local slope of $\frac{\rm{d\omega}}{\rm{dy}}$, which has a
singularity at y= $\frac{3}{4}$L in Model-III, enhances spiral-wave
instability.

\subsection*{Dependence of spiral-wave instability on fibroblast parameters}

\begin{figure}[!ht]
\includegraphics[width=\linewidth]{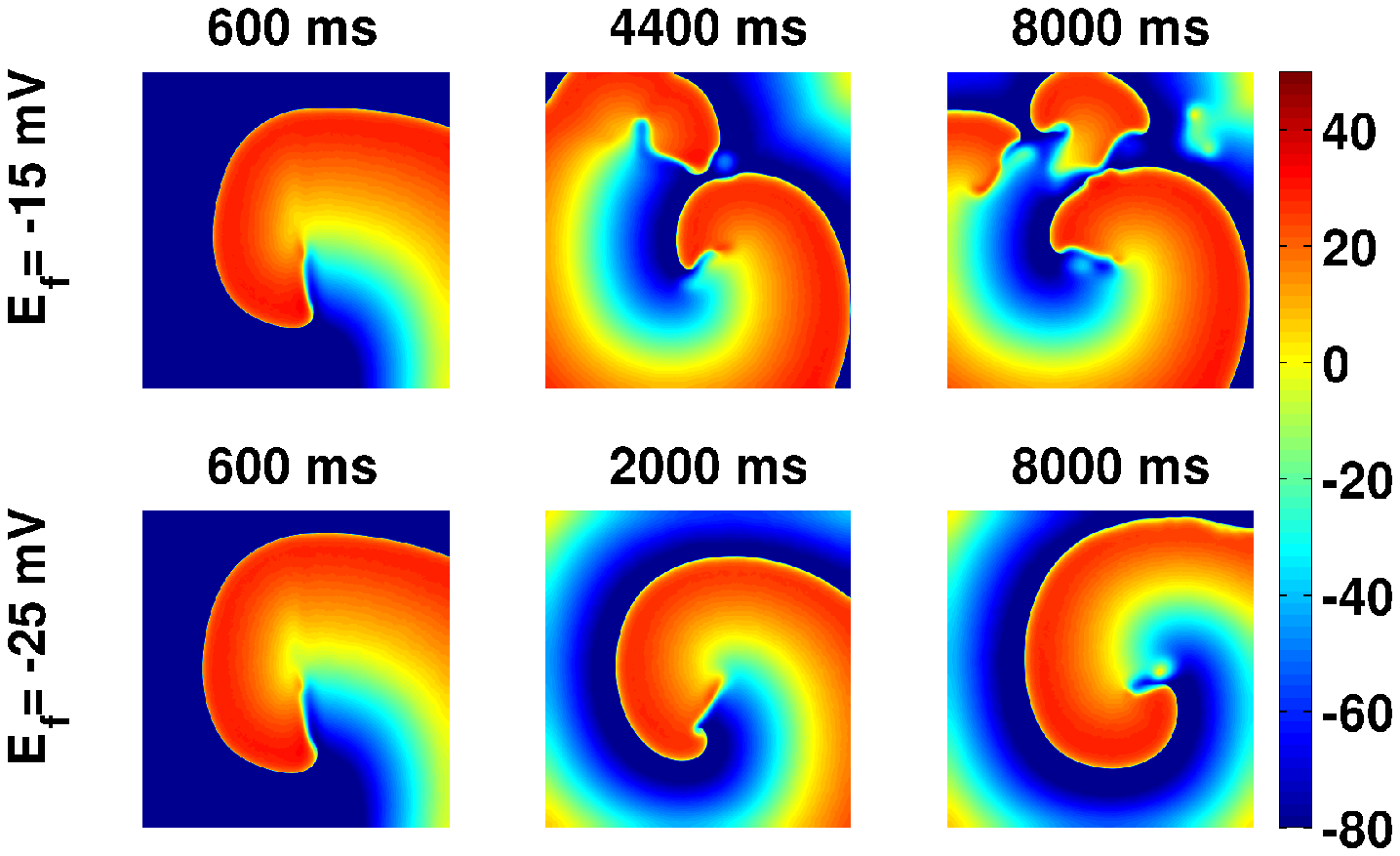}
\caption{ { \bf Dependence of spiral-wave stability on $\rm{E_f}$.} Pseudocolor
plot of $\rm{V_m}$ showing the break-up of a spiral in the presence of GFD,
with the Model-II distribution, $\rm{p_{f1}}$= 10\%, and $\rm{p_{f2}}$=80\% for
$\rm{E_f}$= -15mV (top panel); the same GFD does not lead to break-up for
$\rm{E_f}$= -25 mV (bottom panel). } \label{10to80}
\end{figure}

\begin{figure}[!ht]
\includegraphics[width=\linewidth]{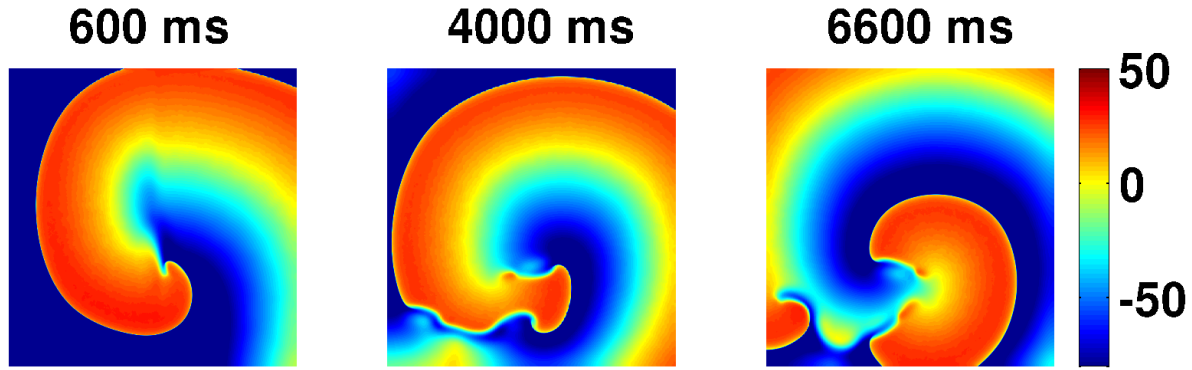}
\caption{ { \bf Spiral-wave break-up in the presence of GFD, with $\rm{E_f}$=
-50 mV.} Pseudocolor plot of $\rm{V_m}$ showing the break-up of a spiral in the
presence of GFD with the Model-II distribution, $\rm{p_{f1}}$=10\%,
$\rm{p_{f2}}$=100\%, and $\rm{E_f}$= -50 mV. The spiral breaks up in the bottom
part of the domain because it is the low-$\omega$ region.} \label{Ef50}
\end{figure}

The APD of a myocyte, in a fibroblast-myocyte composite, depends on $\rm{E_f}$,
and, therefore, the stability of a spiral wave in a domain with GFD also
depends on $\rm{E_f}$. To illustrate this, we first recall the variation of
$\omega$ with $\rm{p_f}$  for three values of $\rm{E_f}$ in figure~\ref{freq},
namely, $\rm{E_f}$= -15 mV (black curve), $\rm{E_f}$= -25 mV (blue curve), and
$\rm{E_f}$= -50 mV (red curve). Note that the change in $\omega$ with
$\rm{p_f}$ is more significant for $\rm{E_f}$= -15 mV than for $\rm{E_f}$= -25
mV. Therefore, in a domain with a given GFD, the spiral breaks up with a lower
GFD for $\rm{E_f}$= -15 mV than for $\rm{E_f}$= -25 mV.  Figure~\ref{10to80}
(top panel) shows the break-up of a spiral wave for GFD with the Model-II
distribution, $\rm{p_{f1}}$= 10\%, $\rm{p_{f2}}$=80\%, and $\rm{E_f}$= -15 mV.
The same GFD does not lead to spiral-wave break-up for $\rm{E_f}$= -25 mV
(figure~\ref{10to80}, bottom panel) (see also the Supplementary Video S3). This
indicates that the readiness with which spiral-wave instability sets in depends
on the degree of variation of $\omega$ induced by the GFD.  For $\rm{E_f}$= -50
mV, $\omega$ decreases with $\rm{p_f}$ (figure~\ref{freq}).  Therefore, in the
presence of GFD, with the Model-II distribution, $\rm{p_{f1}}$= 10\% and
$\rm{p_{f2}}$= 100\%, the spiral breaks up in the bottom region of the domain,
as shown in figure~\ref{Ef50} (see also the Supplementary Video S4). This
demonstrates again that the waves break in the low-$\omega$ region of the
domain.

\begin{figure}[!ht]
\includegraphics[width=\linewidth]{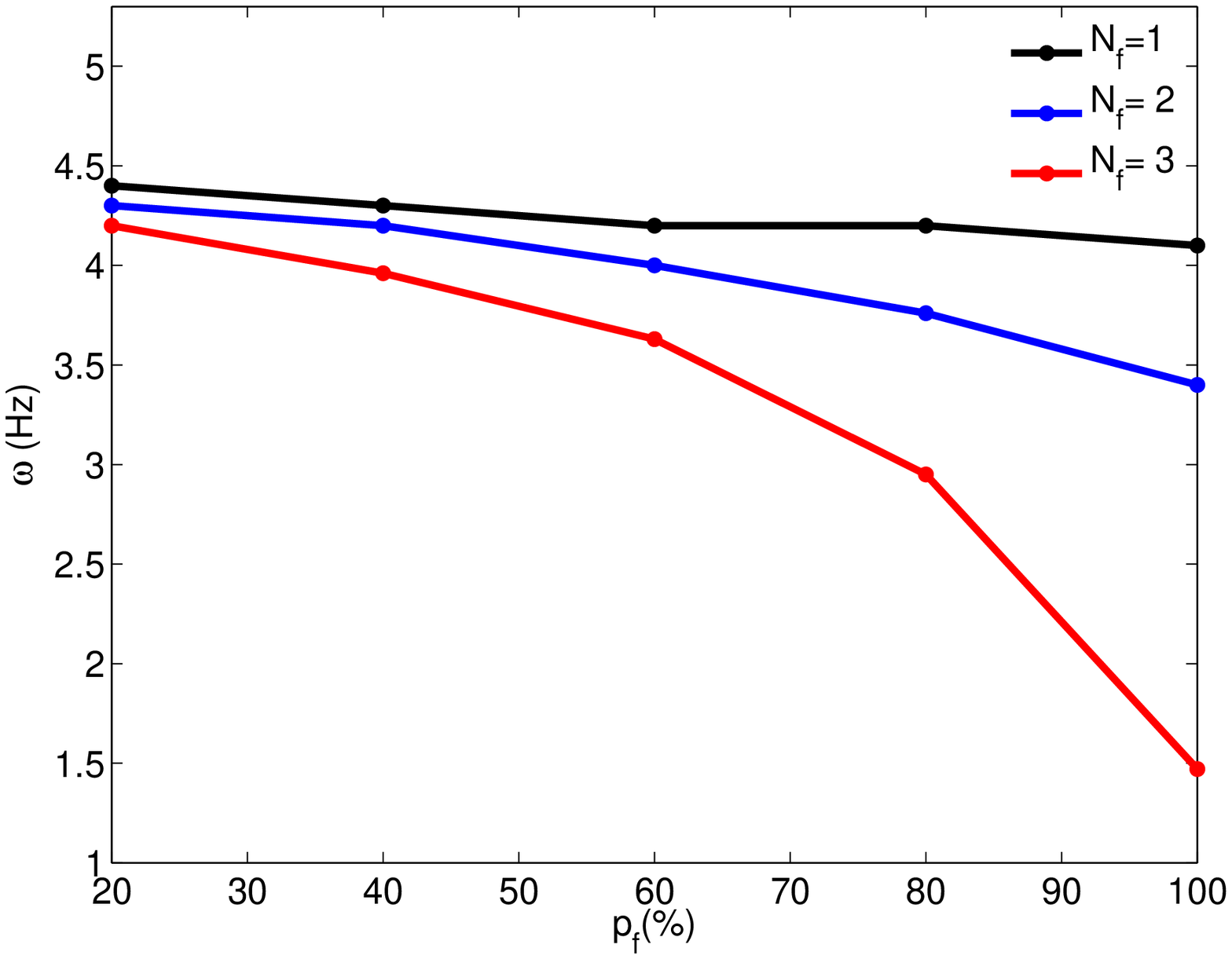}
\caption{ { \bf Variation of the spiral-wave frequency $\omega$ with $\rm{p_f}$
for different values of $\rm{N_f}$.} A plot showing the variation of $\omega$
with $\rm{p_f}$ for three different values of $\rm{N_f}$: $\rm{N_f}$= 1 (black
curve), 2 (blue curve), and 3 (red curve). The value of $\rm{E_f}$ is held
constant at -15 mV for all the curves.  The variation of $\omega$ with
$\rm{p_f}$ is most significant for $\rm{N_f}$= 3.} \label{freqNf}
\end{figure}

\begin{figure}[!ht]
\includegraphics[width=\linewidth]{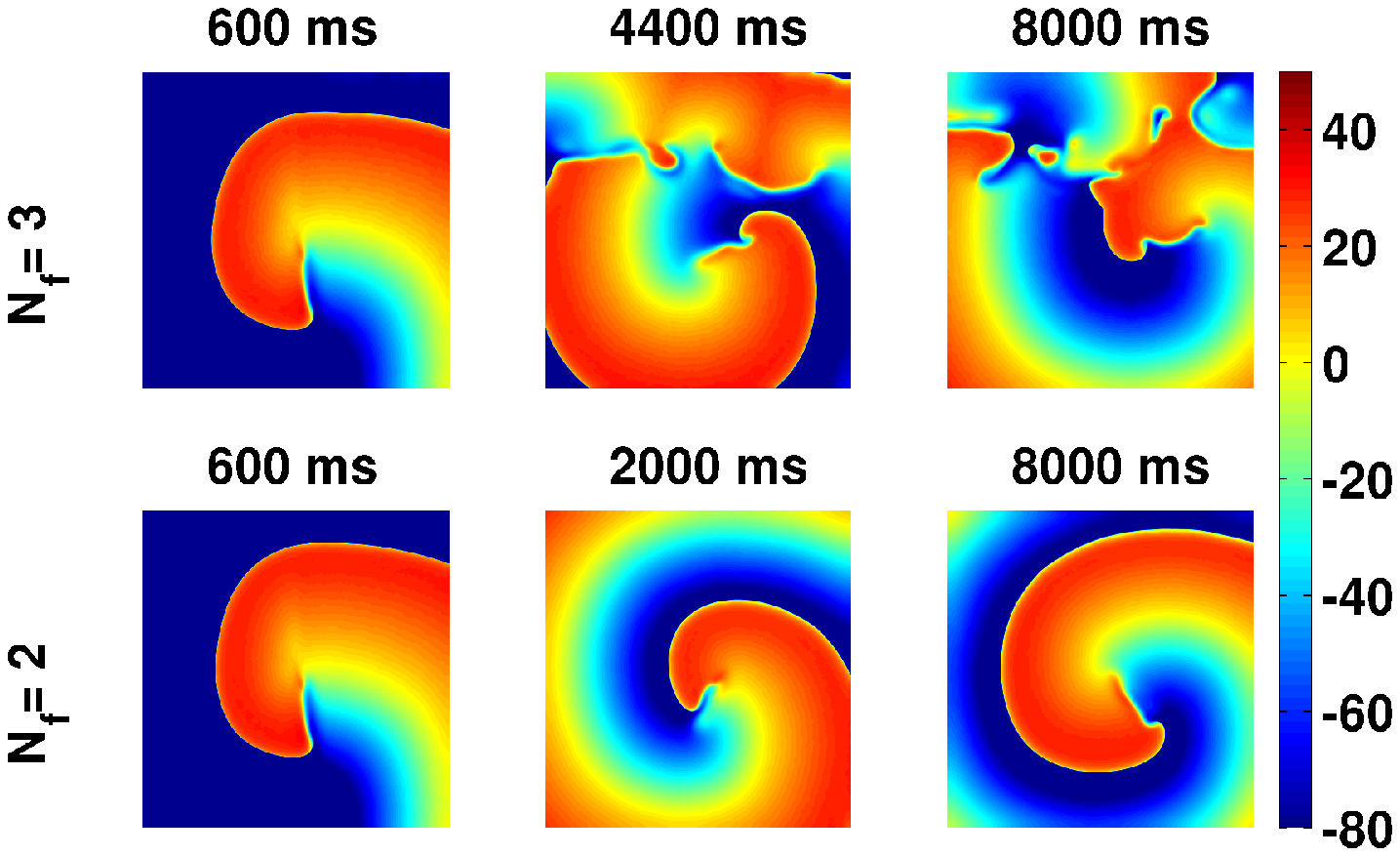}
\caption{ { \bf Dependence of spiral-wave stability on $\rm{N_f}$.} Pseudocolor
plots of $\rm{V_m}$ showing the break-up of a spiral in the presence of GFD with
the Model-II distribution, $\rm{p_{f1}}$=10\%, and $\rm{p_{f2}}$=60\%, for
$\rm{E_f}$= -15 mV and $\rm{N_f}$= 3 (top panel). The same GFD and $\rm{E_f}$
does not lead to break-up for $\rm{N_f}$= 2 (bottom panel).} \label{Nf2_3}
\end{figure}

The variation of $\omega$ with $\rm{p_f}$ also depends on the number $\rm{N_f}$
of fibroblasts that are attached to the myocytes. Figure~\ref{freqNf} shows the
variation of $\omega$ with $\rm{p_f}$ for different values of $\rm{N_f}$, for a
constant value of $\rm{E_f}$= -15 mV. The black, blue, and red curves are for
$\rm{N_f}$= 1, 2, and 3, respectively. These plots show that the larger the
value of $\rm{N_f}$ the larger is the change of $\omega$ with $\rm{p_f}$.
Hence, in a domain with a given GFD and $\rm{E_f}$, the spiral-wave instability
increases with $\rm{N_f}$, because the variation of $\omega$ with $\rm{p_f}$
increases with $\rm{N_f}$. To illustrate this, figure~\ref{Nf2_3} (top panel)
shows the breaking of a spiral wave for a GFD with the Model-II
distribution, $\rm{p_{f1}}$= 10\%, $\rm{p_{f2}}$= 60\%, $\rm{E_f}$= -15 mV, and
$\rm{N_f}$= 3. The same GFD and $\rm{E_f}$ does not lead to spiral break-up for
$\rm{N_f}$= 2 (figure~\ref{Nf2_3}, bottom panel). (See also the Supplementary
Video S5.)

\subsection{Three-dimensional simulation domains}

\begin{figure}[!ht]
\includegraphics[width=\linewidth]{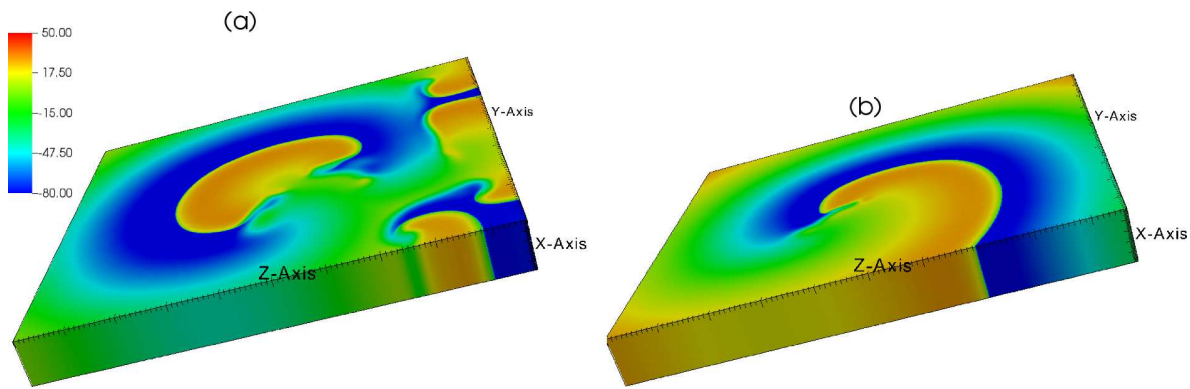}
\caption{ { \bf Dependence of scroll-wave stability on $\rm{N_f}$.} Pseudocolor
plots of $\rm{V_m}$ showing the stability of scroll waves in a 3D domain with
GFD (Model-II distribution, with $\rm{P_f(x,y,z)}$ varying from $\rm{p_{f1}}$=
10\% to $\rm{p_{f2}}$= 60\% along the z-axis and constant in the x-y plane for
a fixed value of z). The value of $\rm{E_f}$ is set to -15 mV.  (a) The scroll
wave breaks up for $\rm{N_f}$=3; (b) the scroll wave remains stable for
$\rm{N_f}$=2.} \label{3d}
\end{figure}

To check if our results also hold in 3D simulation domains, we perform a few
representative 3D simulations, where the thickness (x dimension) is 2 mm and
the linear dimensions in the y and z directions are both 19.2 cm. We initiate a
scroll wave by using the same S1-S2 cross-field protocol that we use in our 2D
simulations for spiral-wave initiation. We find that the results we obtain in
3D, for the dependence of scroll-wave stability on $\rm{E_f}$ and $\rm{N_f}$,
are qualitatively similar to those we have obtained above for 
spiral-wave stability
in 2D. For example, we show the dependence of scroll-wave stability on
$\rm{N_f}$. We take a 3D domaim with GFD given by the Model-II fibroblast
distribution: $\rm{P_f(x,y,z)}$ varies from $\rm{p_{f1}}$= 10\% to
$\rm{p_{f2}}$= 60\% along the z axis, but is constant in the x-y plane for a
fixed value of z; we use $\rm{E_f}=$-15 mV here. Now we study the stability of
a scroll wave for $\rm{N_f}$= 2 and $\rm{N_f}$= 3.  Figure~\ref{3d}(a) shows
a scroll wave breaking up for $\rm{N_f}$= 3; however, the scroll does
not break up for $\rm{N_f}$= 2, as shown in figure~\ref{3d}(b) (see also the
Supplementary Video S6).  Therefore, with GFD, the higher the value of
$\rm{N_f}$ the greater is the instability of scroll waves (as in our 2D
simulations).

\subsection{Pacing-induced re-entry} 

\begin{figure}[!ht]
 \includegraphics[width=\linewidth]{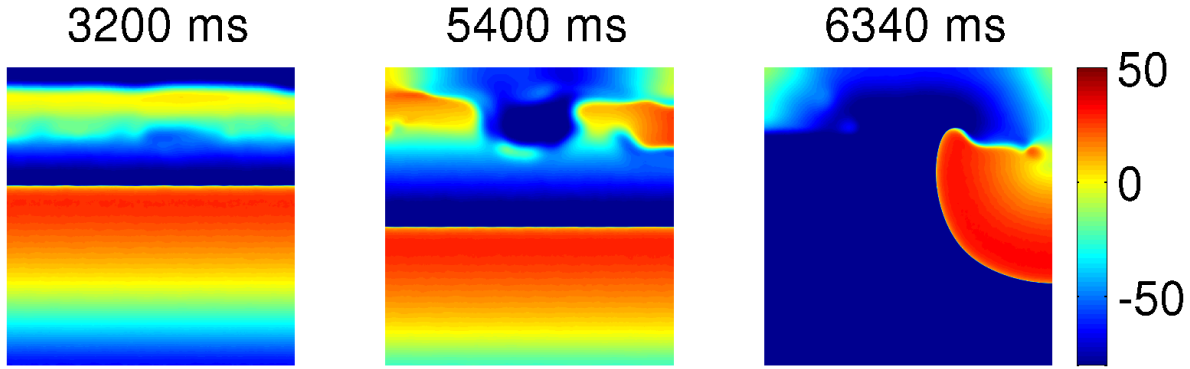}
\caption{ { \bf Pacing-induced spiral wave.} Pseudocolor plot of $\rm{V_m}$
showing the pacing-induced occurrence of a spiral wave, in a 2D simulation
domain with GFD (the Model-II distribution, $\rm{p_{f1}}$= 10\%, and
$\rm{p_{f2}}$= 100\% GFD) paced with PCL= 250 ms. The pacing stimuli are
applied at the bottom of the domain.} \label{PCL250}
\end{figure}

\begin{figure}[!ht]
\includegraphics[width=\linewidth]{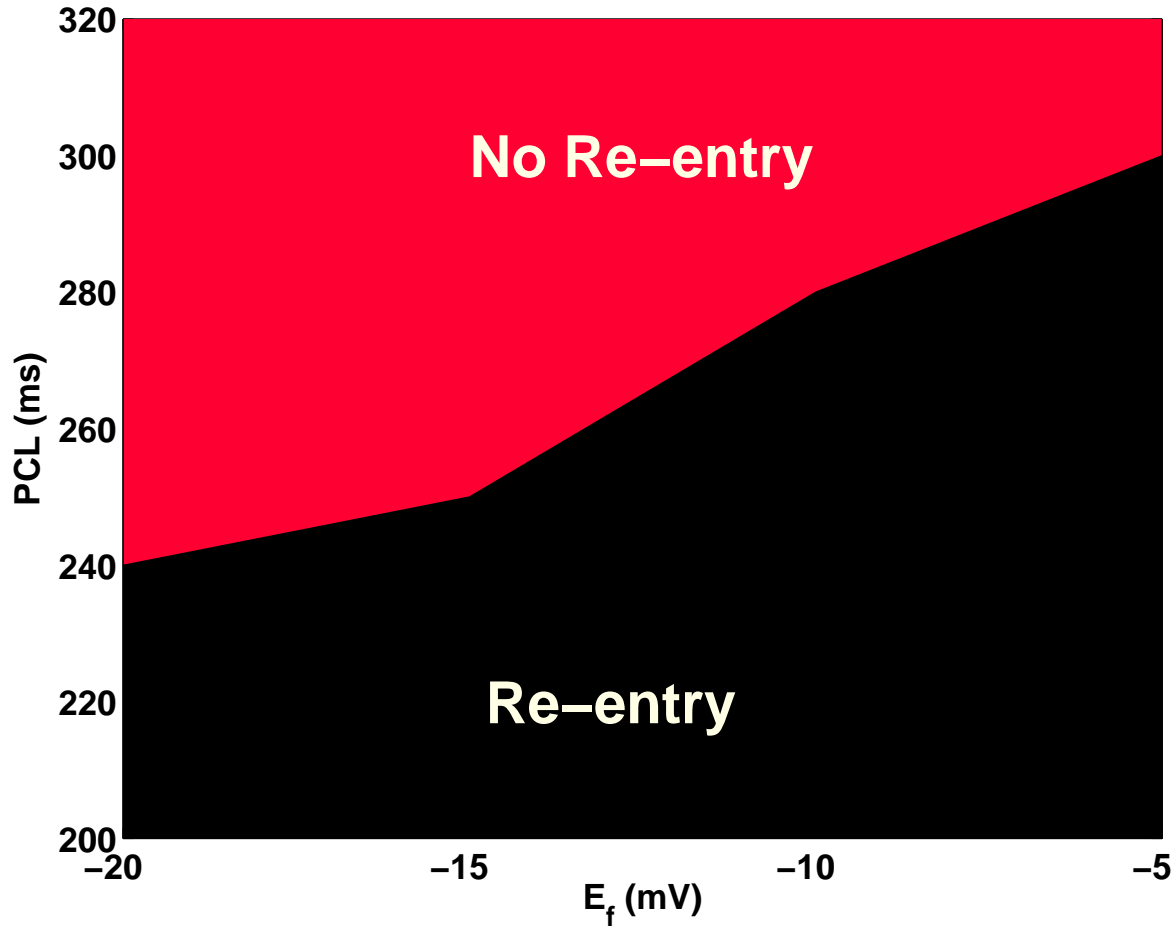}
\caption{ { \bf Stability diagram in the $\rm{E_f}$-PCL plane.} Figure showing
the re-entry (black) and no re-entry (red) regions in the $\rm{E_f}$-PCL plane
(see text). This figure shows that re-entry occurs in the low-PCL region; and,
as $\rm{E_f}$ decreases, the threshold PCL, below which re-entry occurs, also
decreases.} \label{phase}
\end{figure}

So far we have discussed the stability of spiral and scroll waves in the
presence of GFD, where we initiate the spiral or scroll waves by using the
S1-S2 cross-field protocol. We now show GFD itself can lead to re-entry (i.e.,
the formation of spiral or scroll waves) if we pace the system. To illustrate
this, we pace our 2D simulation at a high frequency, with a pacing
cycle length PCL= 250 ms, in the presence of GFD  (the Model-II fibroblast
distribution, $\rm{p_{f1}}$= 10\%, $\rm{p_{f2}}$= 100\%, and $\rm{E_f}$= -15
mV). The pacing stimulus is applied at the bottom edge of the domain.  We apply
20 pulses with this PCL, and we observe that a spiral wave is induced by such
high-frequency pacing, as shown in figure~\ref{PCL250} (see also the
Supplementary Video S7). Thus, GFD not only provides a substrate for spiral- or
scroll- wave instabilities in cardiac tissue, but it can also be a source of
such re-entrant waves if the tissue is paced at sufficiently high frequency.
Such pacing-induced spiral waves occur with high-frequency pacing (low PCL) but
not with low-frequency (large PCL) pacing. In the stability diagram of
figure~\ref{phase}, we show the region in which we observe such pacing-induced
re-entry in the $\rm{E_f}$-PCL plane in the presence of the GFD. The
black-colored region indicates the region in which we obtain spiral waves, and
the red-colored region is the region in which we do not see spiral waves. Thus,
for a given $\rm{E_f}$, re-entry occurs below a certain  PCL; and, as
$\rm{E_f}$ decreases, the threshold of PCL, below which re-entry occurs,
decreases. Re-entry occurs with high-frequency and not with low-frequency
pacing for the following reasons. The wavebacks of the travelling waves
repolarize (come back to the resting state) slowly in the low-$\omega$ regions,
which are in the top part of our simulation domain in figure~\ref{PCL250},
because $\overline{\rm{APD}}$ is large in these regions.  When we use
low-frequency pacing, the wave-fronts do not meet the wavebacks of the
preceding waves. By contrast, when we use high-frequency pacing, the wavefronts
of the succeeding waves interact with the wavebacks of the preceding waves,
and, because of the anisotropic repolarization that arises from the GFD, the
succeeding wavefronts become corrugated, which leads to re-entry and the
formation of spiral waves.         

\section{Discussion}

We have shown by detailed numerical simulations how gradients in the fibroblast
density (GFD) affect spiral- or scroll-wave dynamics in the state-of-the-art
ORd mathematical model for cardiac tissue. Our work shows that such gradients
induce spatial variations in the local spiral- or scroll-wave frequency
$\omega$. These variations play a crucial role in the stability of these spiral
or scroll waves in the presence of GFD. In particular, we find that spiral or
scroll waves break in regions where the local $\omega$ is low. Furthermore, for
a particular GFD, the stability of these waves depend on $\rm{E_f}$ and
$\rm{N_f}$, insofar as they affect the variation of $\omega$ with $\rm{p_f}$
(see figures~\ref{freq} and \ref{freqNf}). Last, but not least, high-frequency
pacing at one end of our domain can lead to the formation of spiral waves if
GFD is present. 
    
Some earlier studies have investigated the effects of fibroblasts on spiral- and
scroll-wave dynamics. Numerical simulations of mathematical models of cardiac
tissue, with a distribution of fibroblasts, have shown single or multiple
spiral or scroll waves \cite{majumder2012nonequilibrium}, depending on the
fibroblast density. A detailed study of the effects of homogeneous or
localized distributions of fibroblasts on spiral-wave dynamics has been carried
out by Nayak, {\it et al.}~\cite{nayak2013spiral}. Fibroblasts, randomly
distributed, have been shown to convert complex wave patterns, like 
mutiple-spiral states, to simpler patterns, like a single rotating spiral state
\cite{nayak2015turbulent}. In a mathematical model of heart-failure tissue,
Gomez, \textit{et al.}, have found that the presence of fibroblasts enhances
re-entrant phenomena~\cite{gomez2014electrophysiological}. Also, the
fibroblast-myocyte coupling has been found to be arrythmogenic, because of
its ability to induce pathological excitations, like
alternans~\cite{xie2009cardiac} and early afterdepolarizations~\cite{morita2009increased}; the latter can, in turn, lead to re-entries.

Although these studies, and many others~\cite{ten2007influence,kazbanov2016effects,nguyen2014cardiac}, have investigated the causes of
re-entrant waves and their sustenance in cardiac tissue with fibroblasts, none
of them has studied, systematically, spiral- and scroll-wave instability
because of the fibroblast-myocyte coupling in a domain with GFD.  Our study, in
which we control GFD, has allowed us to study clearly the arrhythmogenic
potential of such GFD. It is important to carry out such an investigation
because the gradients in density distribution of fibroblasts are known to occur
in real hearts~\cite{morita2009increased}. We hope our results will lead to
detailed studies of GFD-induced spiral- and scroll-wave instability at least in
\textit{in vitro} experiments on cell-cultures. At the simplest level, we suggest
fibroblast analogs of the experiments of Campbell, \textit{et
al.}~\cite{campbell2012spatial}, in which the gradient in $\overline{\rm{APD}}$
has been induced by varying the $\rm{I_{Kr}}$ ion-channel density.

We end our discussion with some limitations of our study. Our tissue
simulations do not take into account anisotropy because of the orientation of
muscle fibers. Such tissue anisotropy may exacerbate the instability of spiral
waves in the presence of a GFD. We have used a monodomain representation of
cardiac tissue; our study needs to be extended to other tissue models, such as
those that use bidomain representations \cite{henriquez1992simulating}. We use
a passive model of the fibroblasts; however, there is some
evidence~\cite{maccannell2007mathematical,li2009characterization,jacquemet2007modelling},
that fibroblasts can behave as active cells. Nonetheless, our qualitative
results, about the arrhythmogenic effects of GFD, and its root cause, should
not depend on such details. 

\section*{Acknowledgements}
We thank the Department of Science and Technology (DST), India, and the Council for Scientific and
Industrial Research (CSIR), India, for financial support, and Supercomputer Education and Research Centre (SERC, IISc) for
computational resources.

\section*{Supplementary Data}

\hspace{1 cm}\textbf{Video S1}. Video showing the break-up of a spiral in the presence of GFD,
with the Model-II distribution, $\rm{p_{f1}}$= 10\%, and $\rm{p_{f2}}$=100\% for $\rm{E_f}$= -25mV. We use 10 frames per
second with each frame separated from the succeeding frame by 20ms in real time.
(MPEG)

\textbf{Video S2}. Video showing the spatiotemporal evolution of a spiral for two cases (A) and (B) of
the fibroblast distribution. The video on the left-panel shows a stable spiral for case
(A), with the Model-II fibroblast distribution, $\rm{p_{f1}}$= 10\%, and
$\rm{p_{f2}}$= 75\%. The right-panel video shows spiral break-up for case (B), with
the Model-III fibroblast distribution, $\rm{p_{f1}}$= 10\%, and $\rm{p_{f2}}$=
65\%. We use 10 frames per second with each frame separated from the succeeding frame by 20ms in real time.
(MPEG)

\textbf{Video S3}. Video showing the break-up of a spiral in the presence of GFD,
with the Model-II distribution, $\rm{p_{f1}}$= 10\%, and $\rm{p_{f2}}$=80\% for
$\rm{E_f}$= -15mV (right panel); the same GFD does not lead to break-up for
$\rm{E_f}$= -25 mV (left panel). We use 10 frames per second with each frame separated from the succeeding frame by 20ms in real time.
(MPEG)

\textbf{Video S4}. Video showing the break-up of a spiral in the
presence of GFD with the Model-II distribution, $\rm{p_{f1}}$=10\%,
$\rm{p_{f2}}$=100\%, and $\rm{E_f}$= -50 mV. The spiral breaks up in the bottom
part of the domain because it is the low-$\omega$ region. We use 10 frames per second with each frame separated from the succeeding frame by 20ms in real time.
(MPEG)

\textbf{Video S5}. Video showing the break-up of a spiral in the presence of GFD with
the Model-II distribution, $\rm{p_{f1}}$=10\%, and $\rm{p_{f2}}$=60\%, for
$\rm{E_f}$= -15 mV and $\rm{N_f}$= 3 (left panel). The same GFD and $\rm{E_f}$
does not lead to break-up for $\rm{N_f}$= 2 (right panel).
We use 10 frames per second with each frame separated from the succeeding frame by 20ms in real time.
(MPEG)

\textbf{Video S6}. Video (bottom panel) showing the break-up of a scroll wave in a 3D domain with GFD (Model-II distribution, with $\rm{P_f(x,y,z)}$ varying from $\rm{p_{f1}}$=
10\% to $\rm{p_{f2}}$= 60\% along the z-axis and constant in the x-y plane for
a fixed value of z), where the value of $\rm{E_f}$ is set to -15 mV and $\rm{N_f}$=3. The same GFD and $\rm{E_f}$ does not lead to break up for $\rm{N_f}$=2 (top panel). 
We use 10 frames per second with each frame separated from the succeeding frame by 20ms in real time.
(MPEG)

\textbf{Video S7}. Video showing the pacing-induced occurrence of a spiral wave, in a 2D simulation
domain with GFD (the Model-II distribution, $\rm{p_{f1}}$= 10\% and
$\rm{p_{f2}}$= 100\% GFD, and $\rm{E_f}$= -25 mV) paced with PCL= 250 ms.
We use 10 frames per second with each frame separated from the succeeding frame by 20ms in real time.
(MPEG)

\end{document}